\documentclass[10pt,twoside]{hsqcd}
\usepackage{epsf,amsmath}
\usepackage{graphicx}

\newcommand{\nutev}{NuTeV}
\newcommand{\ie}{{\it i.e.}}
\newcommand{\eg}{{\it e.g.}}

\begin{document}

\title{SEA QUARK ASYMMETRIES IN THE NUCLEON\footnote{Talk 
given at HSQCD, St Petersburg, Russia, May 2004}\\ A physical model
for parton distributions in hadrons}

\author{J.~ALWALL \\
High Energy Physics, Uppsala University \\ Box 535, 751 21
Uppsala, Sweden\\ E-mail: johan.alwall@tsl.uu.se }

\maketitle

\begin{abstract}
\noindent A physical model for parton densities in hadrons,
based on Gaussian momentum fluctuations of partons and hadronic
baryon-meson fluctuations,
is presented. The model has previously been
shown to describe proton structure function data, and is now applied
to sea quark asymmetries and shown to describe the $\bar d-\bar u$
asymmetry of the proton. By considering fluctuations involving strange
quarks, the model gives an asymmetry between the momentum
distributions of $s$ and $\bar s$, which would reduce the significance
of the \nutev\ anomaly.
\end{abstract}

%\markright{}

%\renewcommand{\@evenhead}

\markboth{\large \sl J.~Alwall  \hspace*{2cm} HSQCD 2004}
{\large \sl \hspace*{1cm} SEA QUARK ASYMMETRIES IN THE NUCLEON}

\section{Introduction} 
Asymmetries of sea quark distributions of the nucleon has for
quite some time been an intriguing problem.  For the part of the
nucleon sea arising from gluon splittings $g\to q\bar{q}$ in
perturbative QCD, symmetry is expected in the distributions of quarks
and antiquarks, \ie\ $q(x)=\bar{q}(x)$, and also $\bar u(x)=\bar
d(x)$. Conventional parameterizations of quark momentum distributions
assume these symmetries also for the $x$-distributions at the start of
the perturbative QCD evolution. However, for these sea distributions
arising from the non-perturbative dynamics of the bound state nucleon
experiment has shown that there are asymmetries between $\bar u$ and
$\bar d$ \cite{Vogt:2000sk}. There are also no symmetry arguments to
prevent asymmetries between quark and anti-quark. This is of great
interest, especially in connection to the \nutev\ anomaly, where the
value of $\sin^2\theta_W$ was found to differ from the Standard Model
fitted value by almost three standard deviations
\cite{Zeller:2001hh}. The anomaly can however, at least in part, be
due to an asymmetry between the momentum distribution of $s$ and $\bar
s$ in the nucleon sea \cite{Davidson:2001ji,Alwall:2004rd}.

Here, we report on recent progress (to be more comprehensively
presented in \cite{AI}) to understand these asymmetries based on a
simple and phenomenologically successful model for the parton
distributions in hadrons. In particular, we have shown \cite{Alwall:2004rd}
that nucleon fluctuations into $|\Lambda K\rangle$, where the $s$
quark is in the heavier $\Lambda$ baryon and the $\bar{s}$ is in the
lighter $K$ meson, gives a harder momentum distribution for the $s$
than the $\bar{s}$. This would reduce the \nutev\ anomaly to
about two standard deviations.

%%%%%%%%%%%%%%%%%%%%
\section*{The model}
%%%%%%%%%%%%%%%%%%%%
\begin{figure}
\includegraphics*[width=30mm]{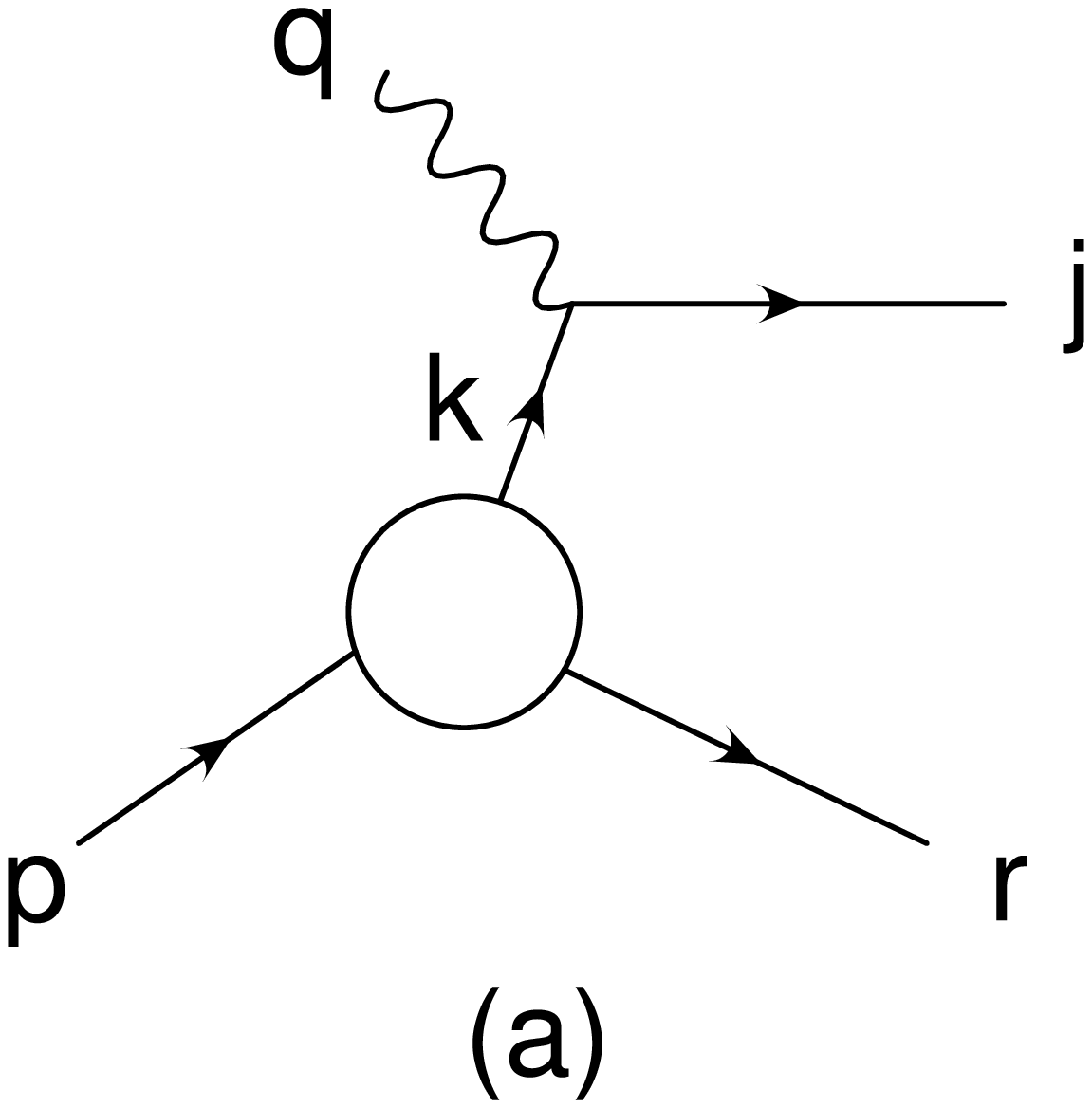}
\includegraphics*[width=30mm]{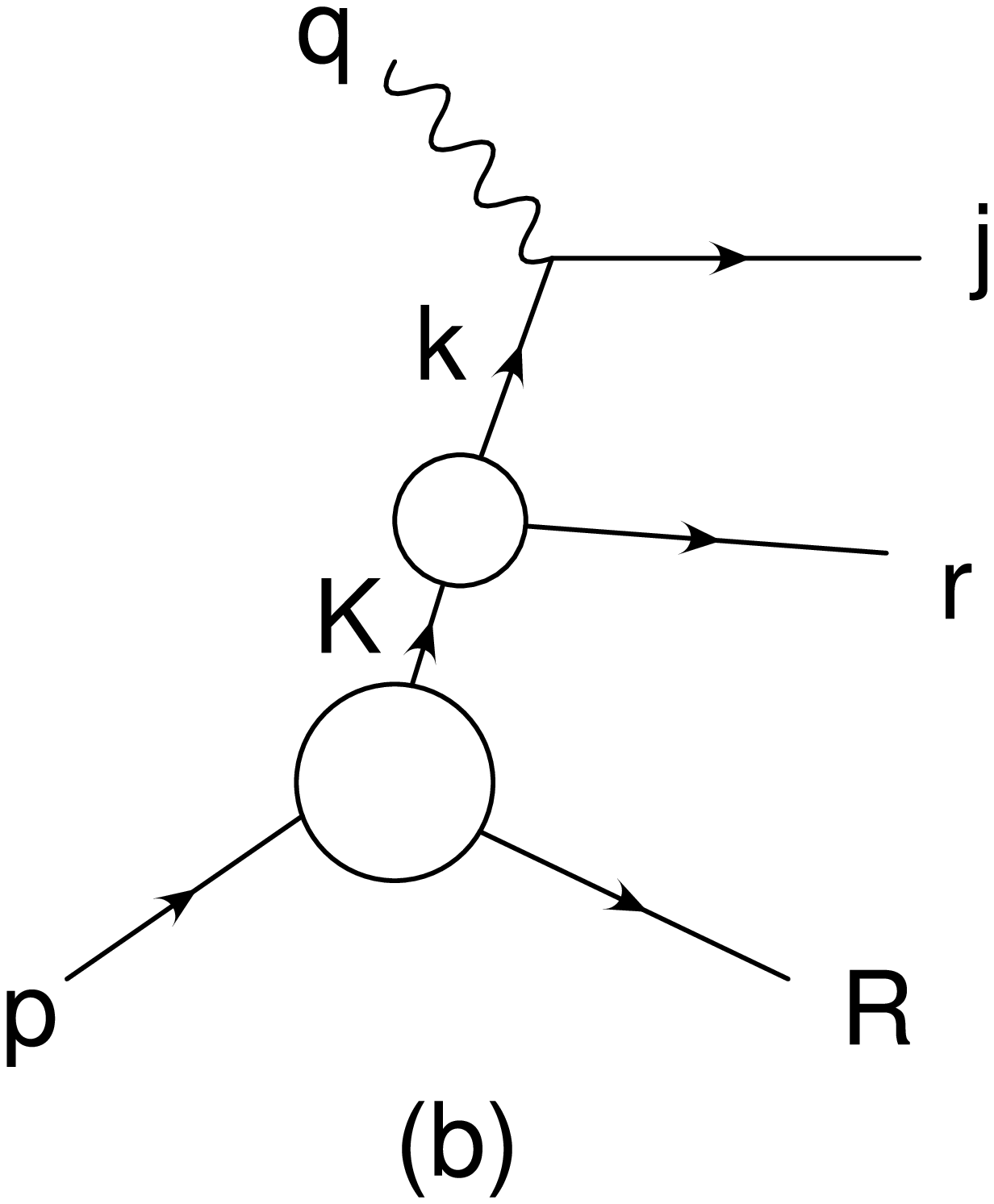}
\includegraphics*[width=60mm]{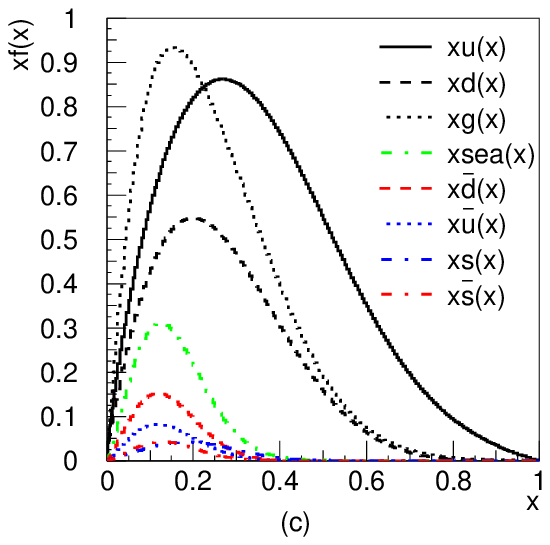}
\vspace*{-5mm}
\caption{\label{fig:fluct} Illustration of the processes (a) probing a
valence parton in the proton and (b) a sea parton in a hadronic
fluctuation (letters are four-momenta). (c) shows the resulting parton
distributions at the starting scale $Q_0^2$.}
\vspace*{-4mm}
\end{figure}
Previously, we have presented a physical model giving the momentum
distributions of partons in the nucleon \cite{Edin:1998dz},
as illustrated in Fig.~\ref{fig:fluct}. The model gives the
four-momentum $k$ of a single probed valence parton by assuming that,
in the nucleon rest frame where there is no preferred direction, the
shape of the momentum distribution for a parton of type $i$ and mass
$m_i$ is then taken as a Gaussian
\begin{equation}
\label{eq:gaussian}
f_i(k) = N(\sigma_i,m_i)
\exp\left\{-\textstyle\frac{(k_0-m_i)^2+k_x^2+k_y^2+k_z^2}{2\sigma_i^2}\right\}
\end{equation}
which may be motivated as a result of the many interactions binding
the parton in the nucleon. The width of the distribution should be of
order hundred MeV from the Heisenberg uncertainty relation applied to
the nucleon size, \ie\ $\sigma_i=1/d_N$. The momentum fraction $x$ of
the parton is then defined as the light-cone fraction $x=k_+/p_+$. In
order to obtain a kinematically allowed final state, we impose the
following constraints. The scattered parton must be on-shell or having
a time-like virtuality, \ie\ have a mass-squared in the range $m_i^2
\le j^2 < W^2$ ($W$ is the invariant mass of the hadronic
system). Furthermore, the hadron remnant $r$ is obtained from
energy-momentum conservation and must have a time-like
virtuality. These constraints also ensure that $0<x<1$.

Using a Monte Carlo method these parton distributions are integrated
numerically without approximations. The normalization of the valence
distributions is provided by the sum rules $\int_0^1 dx\; u_v(x) = 2$
and $\int_0^1 dx\; d_v(x) = 1$, to get the correct quantum numbers of
the proton (and similarly for other hadrons). The gluon normalization
is given by the momentum sum rule $\sum_i \int_0^1 dx \; xf_i(x) = 1$,
where the sum also includes sea partons.

To describe the dynamics of the sea partons, we note that the
appropriate basis for the non-perturbative dynamics of the bound state
nucleon is a hadronic quantum mechanical basis. Therefore we consider
hadronic fluctuations, \eg\ for the proton
\begin{equation} \label{eq-hadronfluctuation}
|p\rangle = \alpha_0|p_0\rangle + \alpha_{p\pi}|p\pi^0\rangle +
\alpha_{n\pi}|n\pi^+\rangle + \ldots + \alpha_{\Lambda K}|\Lambda
K^+\rangle + \ldots
\end{equation}

Probing a parton $i$ in a hadron $H$ of such a fluctuation
(Fig.~\ref{fig:fluct}b) gives a sea parton with light-cone fraction
$x=x_H\, x_i$ of the target proton, \ie\ the sea distributions are
obtained from a convolution of the momentum $K$ of the hadron and the
momentum $k$ of the parton in that hadron. The momentum of the probed
hadron is given by a similar Gaussian as Eq.~\eqref{eq:gaussian} but
with a separate width parameter $\sigma_H$. The kinematical
constraints to be applied in this case are $m_i^2\le j^2 < x_HW^2$ and
that the remnants (see Fig.~\ref{fig:fluct}b) have time-like
virtualities. Here $x_H\sim M_H/(M_\mathrm{baryon}+M_\mathrm{meson})$,
giving a harder spectrum for the heavier baryon than the lighter
meson, for details see~\cite{Alwall:2004rd}. The normalization of the
sea distributions is given by the amplitude coefficients
$\alpha$. These are partly given by Clebsch-Gordan coefficients, but
depend primarily on non-perturbative dynamics that cannot be
calculated from first principles in QCD and are, therefore, taken as
free parameters.

This model results in valence and sea parton $x$-distributions as shown
in Fig.~\ref{fig:fluct}(c). These apply at a low scale $Q_0^2$, and the
distributions at higher $Q^2$ are obtained using perturbative QCD evolution.
The distributions shown in Fig.~\ref{fig:fluct}(c) is the result of
the fits described below.

%%%%%%%%%%%%%%%%%%%%%%%%%%%%%%%%%%%%%%%%%%%%%%%%%%%
\section*{Comparison to data and parameter fitting}
%%%%%%%%%%%%%%%%%%%%%%%%%%%%%%%%%%%%%%%%%%%%%%%%%%%

\begin{figure}
\begin{center}
\includegraphics*[width=110mm]{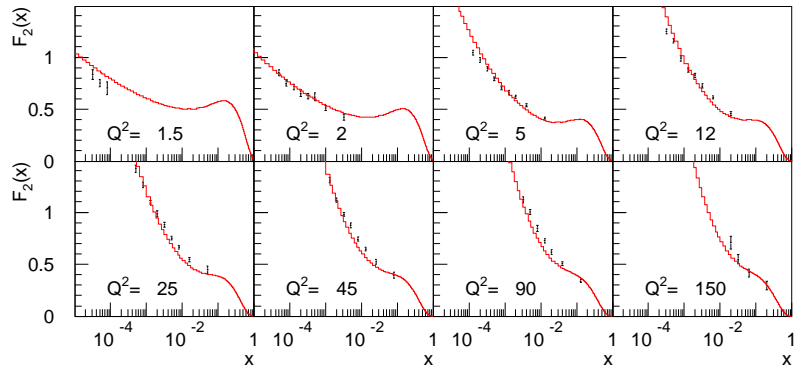}
\end{center}
\vspace*{-10mm}
\caption{\label{fig:f2data} The proton structure function $F_2(x,Q^2)$
from our model compared to HERA H1 data \cite{Adloff:2000qk}.}
\vspace*{-4mm}
\end{figure}

After QCD evolution, the proton structure function $F_2(x,Q^2)$ can
be calculated and the model parameters fitted to data from
deep inelastic scattering. This results in the parameter set
\begin{eqnarray}
\sigma_u &=& 180\, {\rm MeV},\; \sigma_d = 150\, {\rm MeV},\; \sigma_g = 135\, {\rm MeV}
\nonumber \\
\sigma_H &=& 100\, {\rm MeV},\;  \alpha_{sea}^2 = 0.06,\; Q^2_0 = 0.6\,{\rm GeV^2}
\end{eqnarray}
where $\alpha_{sea}^2$ is the fraction of the proton momentum carried
by sea quarks at the scale $Q_0^2$. Such inclusive data can only be used
to determine the overall normalization $\alpha_{N\pi}^2$ for the
dominating light quark sea from fluctuations with pions in
Eq.~(\ref{eq-hadronfluctuation}). The model reproduces $F_2$ data well
in view of the its simplicity, see Fig.~\ref{fig:f2data} and
\cite{Edin:1998dz}.

\begin{figure}
\begin{center}
\includegraphics*[width=90mm]{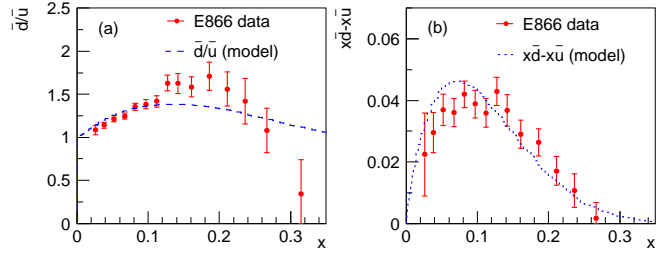}
\end{center}
\vspace*{-10mm}
\caption{\label{fig:dubar} (a) $\bar d(x)/\bar u(x)$ and (b) $x\bar d(x)-x\bar
u(x)$ obtained from the model compared to data from the E866
experiment \cite{Towell:2001nh}.}
\vspace*{-4mm}
\end{figure}

The model also reproduces the observed asymmetry between the $\bar{u}$
and $\bar{d}$ distributions as a result of the suppression of
fluctuations with a $\pi^-$ relative to those with a $\pi^+$, since
the former require a heavier baryon (\eg\ $\Delta^{++}$). With only
one additional parameter the model gives a nice fit to data on this
asymmetry (see Fig.~\ref{fig:dubar}), the parameter in question being
the normalization of the $|n\pi^+\rangle$-fluctuation relative to the
$|p\pi^0\rangle$-fluctuation. The value of this parameter turns out to
be $\alpha^2_{n\pi}/\alpha^2_{p\pi}=1/2$, which might
seem surprising in view of the fact that from isospin the relationship
should be $2:1$. However, these fluctuations implicitly include the
effects of heavier fluctuations like $|\Delta\pi\rangle$. The SU(6)
Clebsch-Gordan coefficient for $|\Delta^{++}\pi^-\rangle$ is much
larger than the coefficient for $|\Delta^{0}\pi^+\rangle$, giving a
larger fraction of $\bar u$ relative to $\bar d$, thus mimicking a
larger $|p\pi^0\rangle$ fraction (see \eg\ \cite{Kumano:1997cy}).

\begin{figure}[b]
\includegraphics*[width=75mm]{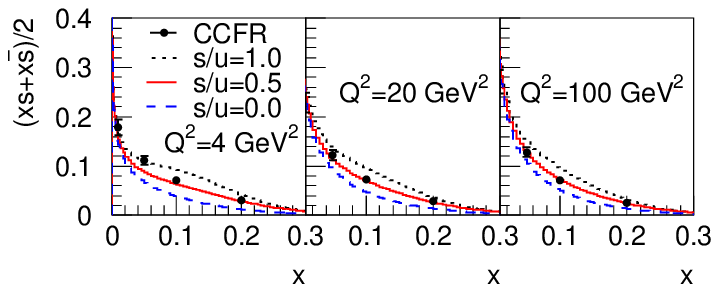}
\includegraphics*[width=60mm]{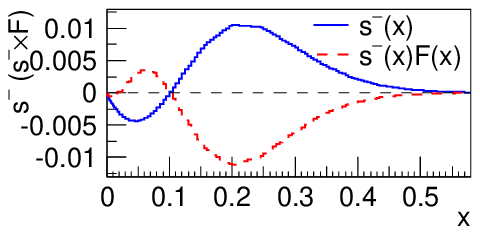}
\vspace*{-10mm}
\caption{\label{fig:CCFR} Left: CCFR deep inelastic scattering data
\cite{Bazarko:1994tt} on the strange sea distribution
$(xs(x)+x\bar{s}(x))/2$ in the nucleon at different $Q^2$ compared to
our model based on $|\Lambda K\rangle$ fluctuations (with
normalization `s/u'$=(s+\bar{s})/(\bar{u}+\bar{d})$) at $Q_0^2$
and evolved to larger $Q^2$ with perturbative QCD that also adds a
symmetric perturbative $s\bar{s}$ component.  Right: The strange sea
asymmetry $s^-(x) = xs(x)-x\bar{s}(x)$ (at $Q^2=20$ GeV$^2$) from the
model and combined with the function $F(x)$ accounting for NuTeV's
analysis \cite{Zeller:2002du}.}
\end{figure}

To fix the normalization of the strange sea in the model, we assume
that all fluctuations including strange quarks (such as $|\Lambda
K^*\rangle$, $|\Sigma K\rangle$) can be implicitly included in the
$|\Lambda K\rangle$-fluctuation and make a fit of the resulting
strange sea ($(s+\bar s)/2$) to data from
CCFR \cite{Bazarko:1994tt}. The result is shown in Fig.~\ref{fig:CCFR}
(left plot), giving a $|\Lambda K\rangle$-normalization
$\alpha_{\Lambda K}/\alpha_{N\pi}\approx 1/5$ such that
$\int_0^1dx(xs(x)+x\bar{s}(x))/\int_0^1dx(x\bar{u}(x)+x\bar{d}(x))
\approx 0.5$, in agreement with standard parameterizations.

%%%%%%%%%%%%%%%%%%%%%%%%%%%%%%%%%%%%%%%%%%%%%%%%%%%%%%%%%%%%%%%
\subsection*{The \nutev\ anomaly and an asymmetric strange sea}
%%%%%%%%%%%%%%%%%%%%%%%%%%%%%%%%%%%%%%%%%%%%%%%%%%%%%%%%%%%%%%%

In the \nutev\ experiment \cite{Zeller:2001hh}, the value of
$\sin^2\theta_W$ was extracted from neutral and charged current
cross-sections of neutrinos and anti-neutrinos. The value they find
differs by about $3\sigma$ from the value obtained in Standard Model
fits to data from other experiments: $\sin^2\theta_W^\mathrm{NuTeV} = 0.2277
\pm 0.0016$ while $\sin^2\theta_W^\mathrm{SM}=0.2227 \pm
0.0004$. A number of possible explanations \cite{Davidson:2001ji} for
this discrepancy have been suggested, both in terms of extensions to
the Standard Model and in terms of effects within the Standard
Model. One explanation of the latter kind would be if the momentum
distributions for $s$-quarks in the nucleon differs from that of $\bar
s$-quarks. In our model, such an asymmetry arise since the $s$ quark
is in the heavier $\Lambda$  baryon and the $\bar{s}$ in
the lighter $K$ meson, giving a harder momentum distribution
for the $s$ than the $\bar{s}$.

In Fig.~\ref{fig:CCFR} (right plot) we show the resulting asymmetry
$s^-(x) = xs(x)-x\bar{s}(x)$ and its combination with a folding
function $F(x)$ provided by NuTeV \cite{Zeller:2002du} to account for
their analysis and give the shift in the extracted value of
$\sin^2\theta_W$. We obtain the integrated asymmetry $S^-=\int_0^1dx\,
s^-(x)=0.00165$, and the shift $\Delta \sin^2\theta_W = \int_0^1 dx\,
s^-(x) F(x)= -0.0017$. Thus, the NuTeV value would be shifted to
0.2260 which is only $2.0 \sigma$ above the Standard Model value,
leaving no strong hint of physics beyond the Standard Model.

\end{document}